\documentclass[preprint]{aastex}

\shorttitle{Secondary Parameters of SNIa Light Curves}
\shortauthors{NSF07-SNIa Collaboration}

\newcommand{\gcm}{g~cm$^{-3}$}
\newcommand{\nni}{\ensuremath{^{56}\mathrm{Ni}}}
\newcommand{\mms}{M\ensuremath{_\mathrm{MS}}}
\newcommand{\mcore}{M\ensuremath{_\mathrm{core}}}
\newcommand{\mch}{M\ensuremath{_\mathrm{Ch}}}
\newcommand{\mref}{M{_\mathrm{ref}}}

\begin{document}

\title{Secondary Parameters of Type Ia Supernova Light Curves}

\author{NSF07-SNIa Collaboration \\ P. H\"oflich\altaffilmark{1},
  K. Krisciunas\altaffilmark{2}, A. M. Khokhlov\altaffilmark{3},
  E. Baron\altaffilmark{4}, 
  G. Folatelli\altaffilmark{5}, M. Hamuy\altaffilmark{5},
  M. M. Phillips\altaffilmark{6}, N. Suntzeff\altaffilmark{2},
  L. Wang\altaffilmark{2} 
  } \altaffiltext{1}{Department of
  Physics, Florida State University, Tallahassee, FL 32306, USA
  pah@astro.physics.fsu.edu} \altaffiltext{2}{George P. and Cynthia
Woods Mitchell Insitute for Fundamental Physics \& Astronomy,
Texas A \& M University, Department of Physics \& Astronomy,
4242 TAMU, College Station, TX 77843, USA;
  krisciunas@physics.tamu.edu, suntzeff@physics.tamu.edu,
  lwang@physics.tamu.edu} \altaffiltext{3}{Department of Astronomy and
  Astrophysics, University of Chicago, Chicago, IL, USA;
  ajk@oddjob.uchicago.edu,vikram@oddjob.uchicago.edu}
\altaffiltext{4}{Homer L.~Dodge Department of Physics and Astronomy,
  University of Oklahoma, 440 W.~Brooks, Rm 100, Norman, OK,
  73019-2061 USA; baron@ou.edu}
\altaffiltext{5}{Departmento de Astronomia, Universidad de Chile,
  Casilla 36D, Santiago, Chile; mhamuy@das.uchile.cl,
  gfolatelli@das.uchile.edu} \altaffiltext{6}{Las Campanas
  Observatory, Casilla 601, La Serena, Chile; mmp@lcoeps1.lco.cl}

\begin{abstract}

High-quality observations of $B$ and $V$ light curves obtained at  Las Campanas Observatory for
local Type Ia Supernovae (SNe~Ia) show  clear evidence that SNe~Ia with
the same brightness decline or stretch may have systematic and
independent deviations at times $t\la 5$ days before and at times $t \ga 30$
days after maximum light. This suggests the existence of two
independent secondary parameters which control the shape of SN~Ia
light curves in addition to the primary light curve parameter, stretch
$s$ or $\Delta m_{15}$.  The secondary parameters may reflect two
independent physical effects caused by variations in the initial
carbon-to-oxygen (C/O) profile in the progenitor
 and the initial central density $\rho_c$
in a carbon-oxygen white dwarf exploding as a SN~Ia. Theoretical light
curves of delayed detonation SN~Ia models with varying progenitor
masses on the main sequence, varying accretion rates, and varying
primordial metallicity 
reproduce two morphologically different and independent types
of variations in observed visual light curves.  These 
calculations predict small variations of $\approx 0.05$ mag in the
absolute brightness of SNe~Ia which are correlated with the variations
of progenitor mass on the main sequence $\mms$ which changes the C/O profile,
and $\rho_c$ which depends on the accretion rate.  Such variations in real
supernovae will induce 
systematic errors in SN~Ia calibration at high redshifts.  A physically
motivated three-parameter, $s$, C/O, $\rho_c$, template for SNe~Ia
light curves might take these variations into account.
Comparison between the theoretical predictions and the observational
results agree qualitatively; however, the observations show variations
between the $B$ and $V$ light curves that are not expected from the
modelling and may indicate limitations in the details of the
theoretical models.

\end{abstract}

\keywords{Supernovae: general}

\section{Introduction}\label{Introduction}

SNe~Ia are thought to be thermonuclear explosions of massive
carbon-oxygen white dwarfs (CO-WD) in binary stellar systems.  These
supernovae are important tools (``standard candles'') of modern
physics and cosmology.  Use of SNe~Ia as standard candles has provided
the first direct evidence of the accelerating expansion of the
universe and the existence of dark energy \citep{riess98,perl99}.

Maximum luminosity varies among SNe~Ia and is not a constant.
Essentially all supernova-based cosmology studies use methods of
calibration of SNe~Ia as standardizable candles (removing luminosity
variations) which rely on empirical relations between the intrinsic
brightness at maximum light and other observable characteristics of
SNe~Ia such as the shape of the light curve, or the rate of brightness
decline after maximum light -- the brightness-decline relation
\citep{p93,phillips99}.  The present accuracy of calibration, $\sim
10$\%, has been sufficient for discovering the Dark Energy, but it
must be improved to perhaps $\simeq 1-2$\% in order to study
properties of Dark Energy quantitatively.  This is a formidable task
which requires increasing the accuracy of SN~Ia light curve
observations, accounting for effects of dust absorption, and so on. It
also requires improving the calibration procedure itself, which
involves two important interconnected issues.

(1) If SN~Ia light curves form a one-dimensional family
characterized 
by a single parameter such as $\Delta m_{15}$,
then the observed spread of individual SNe~Ia around the average
brightness-decline relation can be attributed to random statistical
error, and the accuracy of cosmological measurements can be 
be increased by simply increasing the number of observations of
individual SNe~Ia.  On the other hand, if light curves are
characterized, in addition to $\Delta m_{15}$, by some yet unknown
independent ``secondary'' parameters, then improving the calibration
is impossible without taking the dependence of the light curve on
secondary parameters into account. So far, it is unknown if secondary
parameters exist.  Clear evidence for at least two independent secondary
parameters will be provided in this paper.

(2) Implicit in all SN~Ia calibration procedures is a fundamental
assumption that nearby and distant (cosmological) SNe~Ia behave
identically and that empirical brightness--decline or
brightness--stretch relations established for local SNe~Ia can be used
for high-redshift cosmological supernovae as well. Obviously,
empirical studies of nearby and distant SNe~Ia alone can not confirm
or reject the existence of variations of brightness-decline relations
with redshift.  This requires independent accurate measurements of the
intrinsic brightness of SNe~Ia. Secondary parameters may hold a key to
this difficult problem.  If we understand the physical mechanisms and
the relationship of secondary parameters to initial conditions of SN~Ia
explosions, e.g., metallicity and/or the parameters of binary
progenitors, we may gain some insights into systematic changes in
SN~Ia lightcurves with cosmological time.

The Carnegie Supernova Project has recently obtained a highly uniform
set of SNe~Ia light-curves with an  accuracy of a few
hundredths of a magnitude both for individual SNe~Ia and in terms of
variations between different objects \citep{Con_etal09,Fol_etal09}.
These new data provide clear evidence for the existence of secondary
variations in SNe~Ia light curves which are independent of the primary
Phillips relation, and thus, are evidence for the existence of independent
secondary parameters. The CSP data allow us to begin addressing issues
(1) and (2) outlined above.

SN~Ia models predict a weak dependence of the early light curve
on the carbon-to-oxygen
(C/O) ratio of the progenitor. They also predict a weak dependence of
the  late-time light curve on the initial central   
density, $\rho_c$, of the exploding  WD.
A combined analysis of new observational data and theoretical predictions 
leads us to suggest the existence of two independent secondary parameters 
that, in addition to  $\Delta m_{15}$ or stretch, 
control the intrinsic brightness of SNe~Ia.

The paper is organized as follows:  \S~\ref{sec:previous} gives a
short review of previous work on this subject; \S~\ref{Observations} 
describes new observational data used in this paper; 
\S~\ref{Theory} briefly summarizes the current theoretical
understanding of SNe~Ia and presents theoretical calculations of SNe~Ia
light curves;  \S~\ref{Analysis}  analyzes the  observational light
curves which give the evidence for the existence of secondary
parameters, compares observational and theoretical light curves, and
discusses the theoretical interpretation and the mechanisms by which
secondary parameters arise in SNe~Ia;  results of the paper are
summarized and the discussion of the implications of the results to
the calibration of SNe~Ia is presented in
\S~\ref{DiscussionConclusions}.

\section{Previous Work} \label{sec:previous}

Over the last half decade, a number of observational and theoretical
studies have sought to uncover secondary parameters. Much of this
effort has been in attempts to find direct correlations between
physical effects and peak luminosity.  Examples include metallicity
\citep{wang97,hwt98,timmes03,ellis08,galetal08,howell09,pb08,chamulak08},
asymmetries of the explosion
\citep{wang97b,howell99by01,kasen01el03,kasen04,h06a,kasen09,wwaraa08}, central
density and C/O ratio \citep{hwt98,h00,dominguez01,roepkeetal06,h06},
age of the progenitor \citep{MDVP06,sullivan06}, neutron-rich isotope
to \nni\ ratio \citep{mazzpod06}, and the opacity of the overlying
material \citep{mazzetal01,KW07}. Our approach here is
different. Rather than focusing on a single physical effect on a
specific observable or stage, we make use of detailed stellar
evolution models that were calculated all the way through white dwarf
formation, accretion, and explosion \citep{dominguez01,h06} and focus
on the effects that variations in the progenitor and accretion rate
have on both the early and late parts of the light curve but using
parametrizations when the underlying physics is uncertain.  Both
approaches are valid, and we must take care not to draw conclusions
beyond the range of validity of our {1-D} approach and rely on
consistency checks using observations.  
At the present time, this
approach makes sense since 1-D models do a reasonable job of reproducing
observations. While current 3-D models are able to reproduce some of
the observations, no 3-D model to date has begun with a confirguration
that was the result of detailed stellar evolution,  nor is it
understood what physical variations in 3-D models  reproduce the the tight
brightness-decline relation.
 With or without constant mixing,
1-D models can reproduce the brightness-decline relation and its narrow
width \citep{h96a,umeda99,h02}, a varying amount of mixing produces an
`anti-correlation' and/or a huge spread comparable to the entire range
of SNe~Ia \citep{h96a,pinto,kasen09}.

Though details depend on the pre-conditioning of the WD
\citep{hs02,livne05,kasen09,zingali09}, 3D models of deflagrations and
delayed detonations predict strong mixing of the central region during
the deflagration phase \citep{khokhlov00,Gamezo03,R02} in conflict
with observations of late time spectra \citep{h04,gerardy07} and
remnants \citep{fesen07}.  We note that  recollapsing models
\citep{prd06,prd08,prd09a,prd09b} avoid central mixing, but produce
only \nni\ in the center. In these models the center has expanded
sufficiently prior to carbon ignition that the electron capture rates
have dropped and so burning occurs without neutronization.  The
gravitational confined 
detonation \citep{Plewa04,jordan08,meakin09} model will produce
neutronized material under some ignition conditions and avoid its
production under others \citep{jordan09}.

\section{Observations}\label{Observations}

The light curves of SNe are often stitched together from observations
carried out on a variety of telescopes at a variety of sites.  This has the
advantage of filling in gaps in the light curves.  However, there is a
distinct disadvantage.  The spectral energy distributions of SNe are
considerably different than those of normal stars, and the spectra of SNe
change on time scales of days.  While some spectral features are easily
associated with singly and doubly ionized metals such as silicon and
iron, other absorption features are actually blends of  many lines. 
The effective bandpasses of filters vary from camera to camera.
The net result is that photometry of SNe carried out on different telescopes
often exhibits systematic offsets, sometimes amounting to 0.2 mag.  From
synthetic photometry of spectra of normal stars and spectra of SNe at
different times with respect to maximum light, we can compute
``S-corrections'' which largely resolve these differences
\citep{stritz02,kris03}. 
 However, unless we have good sequences of
spectra for all of our SNe, it is not possible to devise error-free
S-corrections.  Certainly, we wish to attribute variations in light curve
morphology to the SNe themselves, not to some conspiracy of the telescopes,
sites, and cameras.

The Carnegie Supernova Project (CSP), which began operation in September
2004, seeks to address this problem \citep{ham06}. The CSP endeavors to
observe Type Ia and Type II-P SNe in the filters of the Sloan Digital Sky
Survey (u'g'r'i'), the standard Johnson $B$ and $V$ filters, plus the
near-infrared bands $Y$, $J$, and $H$. Roughly 50 SNe are being followed each
year, for five years.  Almost all of the optical photometry of nearby SNe is
being obtained with the Swope 1-m telescope at Las Campanas Observatory
(LCO).  For higher redshift SNe some $BVI$ data are derived from images
with the Wide Field CCD camera on the 2.5-m DuPont telescope at 
LCO, and a
small amount of $BVR$ data is from the 6.5-m Clay Telescope (Magellan \#2).
Based on  high-quality CSP data the RMS deviations of
0.025 mag have been achieved \citep{Con_etal09}.

We use 18 SNe~Ia observed in 2004, 2005, and 2006 as part of the CSP.
All these had well sampled light curves.  Many were observed a week or 
more prior to maximum
light.  Many were observed  60 or more days after maximum light.  For
the purposes of this paper we restrict ourselves to the $B-$ and, in particular, $V-$band
photometry because, based on theoretical models, $V$ is expected to be least sensitive
to additional variables.

The $BVRI$ templates of SNe~Ia are provided in \citet{prieto06}.
Sets of $BVRI$ templates are characterized by the standard decline
rate parameter  
$\Delta m_{15}(B)$
which serves  as a morphological label for the set. While
\citet{prieto06} provide $BVRI$ templates only $BV$ were used in this
analysis. 
Prior to fitting the light curves with the templates of
\citet{prieto06} we first 
estimated the time of $B$-band maximum and subtracted the redshift- and
time-dependent K-corrections from the photometry by interpolating the $B$- and
$V$-band corrections of \citet{hamuy93}.  If the subsequently determined
time of maximum light was more than 0.5 days different than the value adopted for
the calculation of the K-corrections, then they were
recalculated. It was then checked that the newly determined time of
maximum light was statistically consistent and thus there was no need
to iterate further. 

The range of $\Delta m_{15}(B)$ for the \citet{prieto06}
templates is $0.83 - 1.93$. For example, for a given object we chose
$B$- and $V$-band templates from \citet{prieto06}, stretched them by
the time dilation factor (1+$z$), then shifted them by small
increments over a range of dates and over a range of magnitudes to
minimize the total $\chi ^2$ of the fit in each filter.  Then we tried 
the other templates including all bands over a range of $\Delta m_{15}(B)$ to determine
which templates give the lowest $\chi^2$ of all. In this way we
determined the observed maximum magnitudes and the times of maximum
light.  For each of our SNe we obtained functions that fit the $B$-
and $V$-band light curves from roughly -5 until 25 days after the time of
maximum light.
 This time interval for fitting the data was motivated by the theoretical
models  which predicted  that, during this period, the visual LCs should be
least effected by variations in the central density and the
progenitors \citep{hwt98,dominguez00,h06}.

For a comparison of the fits to the photometry, we then subtracted off the
derived maximum magnitudes in $B$ and $V$, subtracted off the times of
maximum light in the two filters, and divided the ``time-since-maximum'' by
(1+$z$) to give ``rest frame days since maximum''.  The same normalization is
applied to the photometric data.

\section{Models of SNIa}\label{Theory}

 \subsection{Explosion}

Current observations of SNe~Ia favor a delayed detonation (DD) scenario
of a SN~Ia explosion  in which the explosion begins as
a subsonic deflagration which later turns into a supersonic detonation
by the process of a deflagration-to-detonation transition or
DDT \citep{khokhlov91}. The ensuing detonation 
incinerates the entire WD.  In one-dimensional models the deflagration
speed, $S$, and the transition density, $\rho_{tr}$, at which the DDT
occurs are free parameters.  The value of $\rho_{tr}$ determines the
fraction of a WD that will burn to nuclear statistical equilibrium
and produce \nni. Therefore, $\rho_{tr}$ is a critical
parameter that controls the brightness of a SNe~Ia. DD models with
$\rho_{tr} \simeq 0.5 - 2.5\times 10^7$~\gcm\  reproduce the observed
range of SNe~Ia luminosities, correct stratification of chemical
elements in SNe~Ia envelopes, and the correlation between maximum
brightness and width of the light curve consistent with observations
\citep{h95,h96a,Mazzali1998,mazzetal01,h02,h06,quimby07,mazzali07,kasen09,marion09}.
  DD models appear to be in agreement with
observations of SN~Ia remnants \citep{fesen07,badenes08}.

The initial central density of a WD, $\rho_c$, its metallicity, and its C/O
ratio also influence the production of \nni\ and maximum
brightness of SNe~Ia during the explosion.
Note that in our models, the free parameters are the main sequence
mass of the progenitor, $\mms$, the accretion rate, 
and the primordial metallicity $Z$ \citep{dominguez01,h06}.
The fact that the brightness of a 
SN~Ia is controlled, for the most part, by a single, primary
parameter $\rho_{tr}$, and that there exists an evolutionary bottleneck
associated with the limiting Chandrasekhar mass of a WD provides a
plausible explanation for a near one-dimensional sequence of SN~Ia
explosions.  A crucial point is that $\rho_c$, primordial metallicity, and C/O
ratio influence certain characteristics of the explosion \textsl{not
associated} with the primary brightness-decline relation. In
particular, the C/O ratio has an influence on the expansion velocity
of SNe~Ia.  The dependence is caused by variations in nuclear
binding energy of the CO fuel. The larger the C/O ratio, the smaller
the binding energy, and the faster the SN~Ia envelope expands. This
is a secondary effect since the bulk of the kinetic energy is
determined by a much larger difference in binding energies of C/O and
products of explosive burning (Fe-peak elements). On the other hand,
$\rho_{c}$ influences the distribution and amount of \nni\ in the
innermost parts of a SN~Ia.  SN~Ia models based on explosions of
Chandrasekhar-mass WDs predict a hole in the \nni\ distribution near
the center which is filled instead with highly neutronized isotopes of
Fe-group elements. The hole is caused by electron captures and
neutronization of matter at high densities.  The size of the hole
increases with $\rho_{c}$. These two effects influence the formation
of both early and late portions of a light curve. As noted above the
nickel hole is generally absent in all 3-D models, either because the
core expands prior to ignition, or because in delayed detonation
models mixing occurs during the deflagration phase, but the stratified
structure is restored by the detonation. Because the flame is very
topologically complex during the deflagration phase, the detonation
will not produce a central neutronized hole.

The initial C/O ratio and profile and $\rho_c$ in a supernova are a
result of stellar and binary evolution. The C/O profile is produced
during the central Helium burning and thin shell burning during the
stellar evolution and the accretion to $\mch$. The central He burning
is initially dominated by carbon production via the 3$\alpha$ reaction
in the convective core. When the He mass fraction becomes depleted,
$^{12}C(\alpha,\gamma)^{16}O$ mainly controls He-burning and most of
the $^{16}O$ is produced during the late phases of central
He-burning. Note that the final abundances depend on a combination of
the $^{12}C(\alpha,\gamma)^{16}O$ rate and chemical mixing which
determines the duration of the phase of depleted He-core burning.  
Stellar evolution
models which produce low C/O ratios of $\simeq 0.25-0.4$ are in
agreement with observational constraints, like the amount of oxygen
found in the inner zone of pulsating white dwarfs or the age of open
clusters \citep{dominguez99,dominguez01,metcalfe01,straniero03}.  The
C/O ratio in the burning shell is greater, 
$\approx 1$, because the shell helium source has lower density and
higher temperature compared to helium burning in the core.  The size
of the convective core depends mainly on the progenitor mass \mms\ on
the main sequence and, to some extent, on the primordial metallicity
$Z$, namely the iron abundance, which dominates the opacity
\citep{dominguez01,h00} The dependence of \mcore\ on \mms\ is
nonlinear, with the mass of the He convective core changing with \mms\
slowly for low \mms\ and 
more rapidly when \mms\ approaches its maximum value. Approximately,
\mcore\ is $0.3,0.4, 0.7M_\odot$ for $\mms = 1.5,5,7 M_\odot$,
respectively \citep{dominguez01}. The core's C/O ratio is smallest for
small \mms, increases rapidly when \mms\ increases from $1.5$ to $3
M_\odot$ and remains relatively constant for larger \mms. These two
effects combine to make the mean C/O ratio in an exploding WD decrease
with \mms\ slowly for low mass stars $\mms \simeq 1.5 - 5M_\odot$ and
much more rapidly for higher mass stars with $\mms \simeq 5 - 7
M_\odot$.  In turn, this makes the explosion energy of a WD more
sensitive to \mms\ when \mms\ is large.

The mass of the WD core is sensitive to the mass of the progenitor on
the main sequence and is rather insensitive to the primordial
metallicity. To first order, the C/O ratio and thus,
the explosion energy decreases with $\mms$. This tendency is
generally true for $\mms $ larger than $3 M_\odot$ independent of 
assumptions about mixing and nuclear reaction rates.  For lower mass
progenitors, the total mean C/O ratio varies little with
$\mms$. $\mcore$ decreases with $\mms$, but this effect is almost
compensated by a decrease of the local C/O ratio which decreases with
$\mms$.  Both this effect and the high C/O ratio in shell burning
can be understood by nuclear physics and chemical mixing.  The local
C/O ratio depends mostly on the competition between triple--$\alpha$, $3\,
 ^{4}He \rightarrow ^{12}C$, and $\alpha$-capture on $^{12}$C,
i.e. $^{12}$C$(\alpha,\gamma)^{16}$O.  As a three-body reaction,
triple--$\alpha$ dominates at high $\alpha$ concentrations whereas
$^{12}C(\alpha,\gamma)^{16}O$ 
dominates when $^4He$ is depleted (and at high temperature). During the
early phases of helium core burning mostly $^{12}$C is produced but,
eventually it is depleted by $^{12}$C$(\alpha, \gamma ) ^{16}$O. For
burning in small convective cores,
even moderate chemical mixing keeps $^4$He at a certain
level and prolongs the phase of $^4$He burning under depleted
conditions,  reducing carbon. In burning in thin
radiative shells, the temperatures are higher and the 
burning time-scales are shorter, therefore less carbon is depleted.

The central density $\rho_c$ at which the WD ignites is controlled by
the competition between adiabatic compression caused by accumulation
of mass at the surface and energy losses in the center of the WD. 
As a result, $\rho_c$ depends on a 
configuration of the binary system, that is on the rate of accretion
onto the WD from the stellar companion.

The
physics of the ignition process is multi-D in nature.  With the exception
of \citet{gw95} who find ignition occurs in rising plumes, all
multi-D simulations to date show ignition at or
near the center due to the downward motion of plumes
\citep{hs02,zingali09} in the simmering phase.  Though relevant for
the preconditioning of 
the explosion, variations in the final stage of runaway are of the
order of hours \citep{hs02} and unlikely to effect the central
density.  As a result, $\rho_c$ depends on the configuration of the
binary system, the evolutionary state of the stellar companion, and
the resulting accretion rate.

In what follows, we characterize the initial conditions by $\rho_c$,
and the progenitor characterized by the initial metallicity Z and
\mms. We use initial distributions of C and O in the core as predicted
by the evolutionary calculations of a main sequence star with appropriate
\mms. Both during stellar shell burning and He burning during the
accretion  $C/O\simeq 1$.  Once the
stellar and WD 
evolution has been calculated to the onset of the explosion, the
explosion is calculated using one-dimensional DD models taking
into account the progenitor evolution, the hydrodynamics of the
explosion, detailed nuclear networks with 213 isotopes, the radiation
transport, detailed atomic models, and $\gamma $-ray transport.
Details of the actual models are described in \citet{dominguez01} and
\citet{h06}.

\citet{dominguez01} described the effects of varying the
$^{12}$C$(\alpha, \gamma ) ^{16}$O rate from the high value
\citep{cfhz85} used in our calculations to the lower value of
\citet{cf88}.  The effect on the final
compositions is large, but we use the value that has been shown to
agree with observational constraints
\citep{dominguez99,dominguez01,metcalfe01}
Note, however, that the final C/O ratio depends  on
both the rate and the chemical mixing scheme adopted in the stellar
models \citep[see for example][]{straniero03}.

\subsection{Theoretical light curves \label{sec:lc}}

We calculated $B$ and $V$ light curves of a series of DD models with
fixed $\rho_{tr} = 2.3\times 10^7$~\gcm\ which produce explosions with
spectral and light curve characteristics of normal bright SNe~Ia.  Our
fiducial model has $\rho_c=2\times 10^9$~\gcm, primordial metallicity
$Z$ equal to the solar value $Z_\odot$ \citep{ag89}, and $\mms = 5
M_\odot$. The abundances in the WD are a result of the stellar
evolution of a main sequence star with metallicity, $Z_\odot$.  When
scaling $Z$, for elements up to Si, we adopted the [0/Fe] abundance
suggested by \citet{argast} which implies smaller variations with
redshift for elements up to Si compared with Fe by a factor of three.
This is done because $^{22}$Ne affects the explosive nucleosynthesis
whereas Fe determines the opacity, and therefore, the size of the
convective He-burning core (for a given mass) and, to some extent, the
$B$-band magnitudes. Note that the effect of primordial metallicity
$Z$ on the explosive nucleosynthesis is dominated by the $^{22}$Ne
abundance and not by the iron abundance because it is the reduction of
the proton/nucleon
ratio, $Y_e$, which changes the 
explosive equilibrium abundances.  The light curves presented below illustrate
the effects that variations of central density and progenitor mass,
\mms, have on light curves of SNe~Ia.

\subsubsection{Influence of progenitor mass \mms\  and metallicity}

\begin{figure}
\epsscale{.80}
\plotone{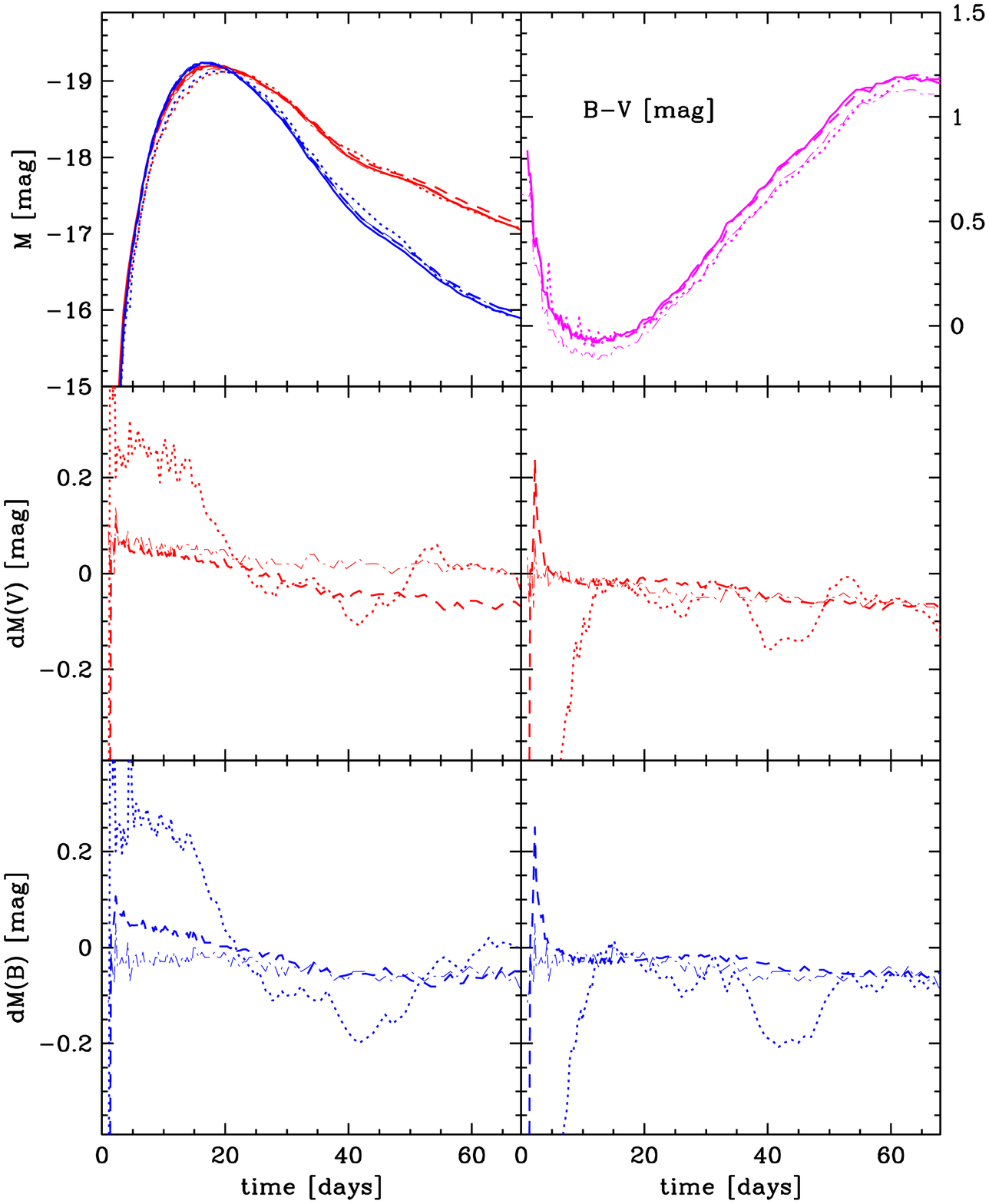}
\caption{$B$ (blue) and $V$ (red) and $B-V$ are given for a number of
  delayed detonation 
models with the same explosion parameters  \citep{dominguez01} and
$\rho_c = 2\times 
10^9$~\gcm\ (see \S~\ref{Theory}) but varying 
progenitor mass \mms\ between $1.5$ and $7.0M_\odot$ and metallicities
$Z$ between 
$0.002$ and $0.02$ (solar). Models are referenced by the pair of numbers
$[\mms,Z]$ 
for $[1.5,0.02]$ (dashed), $[7.0,0.02]$ (dotted) and $[5.0,0.002]$
(dash-dotted). 
The reference model has $\mms  = 5 M_\odot$, and $Z=0.02$ (solid).
The $B$ (blue) and $V$ (red) magnitudes and the color index $B-V$ are
given in the upper right and 
left panel respectively.                                                      The $\Delta m_{15}$ for both $V$ and $B$ light curves are close to
within $0.03$ mag but they are not 
identical. The lower panels show the $B$ and $V$ differentials without
and with stretch-correction to the 
$s_f$ of the reference model on the left and right, respectively (see
Table 1). 
\label{prog}} 
\end{figure}

Fig.~\ref{prog} presents $B$ and $V$ light curves and $B-V$ of four DD
 models with fixed $\rho_c=2\times 10^9$~\gcm\ and varying \mms\ and
 primordial metallicity (upper panels). The lower panels present a
 differential comparison of light curves normalized to maximum light.
 The difference is defined as $dM(t) = M(t) - M_f(t)$, where $M_f$ is
 the magnitude of the fiducial DD model.  The figure shows that
 variations in the progenitor, i.e. the main-sequence mass \mms\ and
 metallicity, strongly change the rise to maximum light.  These
 variations are caused by variations in expansion velocity in models
 with various \mms\ and hence with various C/O ratios.  The expansion
 velocity decreases when \mms\ increases and the overall C/O ratio
 decreases.  At the same time, the chemical and density structure of
 the outer parts of the SN~Ia envelope is similar for all DD
 models. Therefore, variations in the formation of the early
 light curve are mostly controlled by the rate at which the outer
 layers expand and become transparent. The faster the expansion rate,
 the faster the photosphere recedes, and the faster the light curve
 rises towards maximum (see the upper left panel of Fig.~\ref{prog}).
 The light curve of the $\mms=7 M_\odot$ model in Fig.~\ref{prog}
 rises notably slower than the fiducial light curve while the light
 curve of $\mms = 1.5 M_\odot$ model rises somewhat faster. This is
 reflected as the early time negative and positive $dM$ in the left
 panels of Fig.~\ref{prog}. The effect is more pronounced for higher
 \mms\ due to the greater sensitivity of \mcore\ on \mms\ near the
 high end of \mms\ interval. In particular, a strong secondary
 extremum develops when \mcore\ extends to layers which only undergo
 incomplete silicon burning.  As a consequence, we expect that
 differential effects are most pronounced in stellar populations with
 a mix of young and old stars, since these will contain a distribution
 of \mms.  We notice also that variations of the
 progenitor lead to small variations on a 10 percent level in $B-V$ and
 its evolution with time (upper right panel).  Changing the
 primordial $Z$ will increase the primordial iron abundance in the outer
 layers, and, at the expense of $^{56}$Ni, more $^{54}$Fe will
 be produced from $^{22}$Ne.  Primordial metallicity plays a minor role for
 variations in $V$, but $B-V$ becomes bluer with decreasing primordial
 metallicity. 
 This direct `photospheric' effect does not change the visual LCs but
 the B and UV light curves \citep{hwt98}.  The evolution in $B-V$ is
 similar to and has been discussed in \citet{kris03}.

If we were to ``observe'' these four supernovae, match their light
 curves by applying the stretch correction, and \textsl{then} compare the
 residual differences we would get the result shown in the lower right
 panels of Fig.~\ref{prog}.  We show $dM(t)$ after matching the light
 curves around maximum and determining the stretch over the range up to
 $\simeq 15$ days past maximum light. The end result is essentially
 identical light curves near and past maximum with the exception of a
 strong deviation in the rise time for the light curve with the
 highest $\mms=7 M_\odot$ and a small deviation of the same light
 curve approximately $30$ days past maximum. The pre-maximum deviation
 of the $\mms =1.5 M_\odot$ light curve is smaller but still visible
 at the level of $\simeq 0.1$ mag.

There has long been a suspicion that the metallicity of the progenitor
should be associated with the luminosity at peak
\citep{wang01,hametal95,blue96}.  Recently attempts have been made to
measure directly the average metallicity in the environment of the SN
using either line indices or ratios
\citep{hametal00,galetal05,galetal08} or by measuring the spectral
energy distribution (SED) of the
galaxy \citep{howell09}. These studies haved failed to confirm the
expected shift in peak luminosity with metallicity.
Nuclear physics predicts that primarily due
to increased amounts of the neutron rich $^{22}$Ne there should be a
direct correlation between the amount of \nni\ produced and the
progenitor metallicity \citep{timmes03}. Recent results
\citep{howell09,Neill10} find a weaker than expected dependence of
metallicity on  \nni,
whereas \citet{galetal08}
report
a metallicity dependence in the Hubble residual, but not the peak luminosity.
 From the results of
\citet{argast} the variation of elements below Si (including Ne)
varies less than Fe by about a factor of three, thus the sensitivity
of the peak brightness is smaller than would be found by just scaling
all elements to the iron abundance. Theoretical
attempts to find correlations between metallicity and peak luminosity
(or light curve shape) have not followed the detailed stellar
evolution through 
to explosion, but have rather just altered the nucleosynthetic yields
post-explosion and calculated the light curve, scaling just on the
iron abundance
\citep{kasen09}.

\subsubsection{Influence of central density $\rho_c$}

\begin{figure}
\epsscale{.80}
\plotone{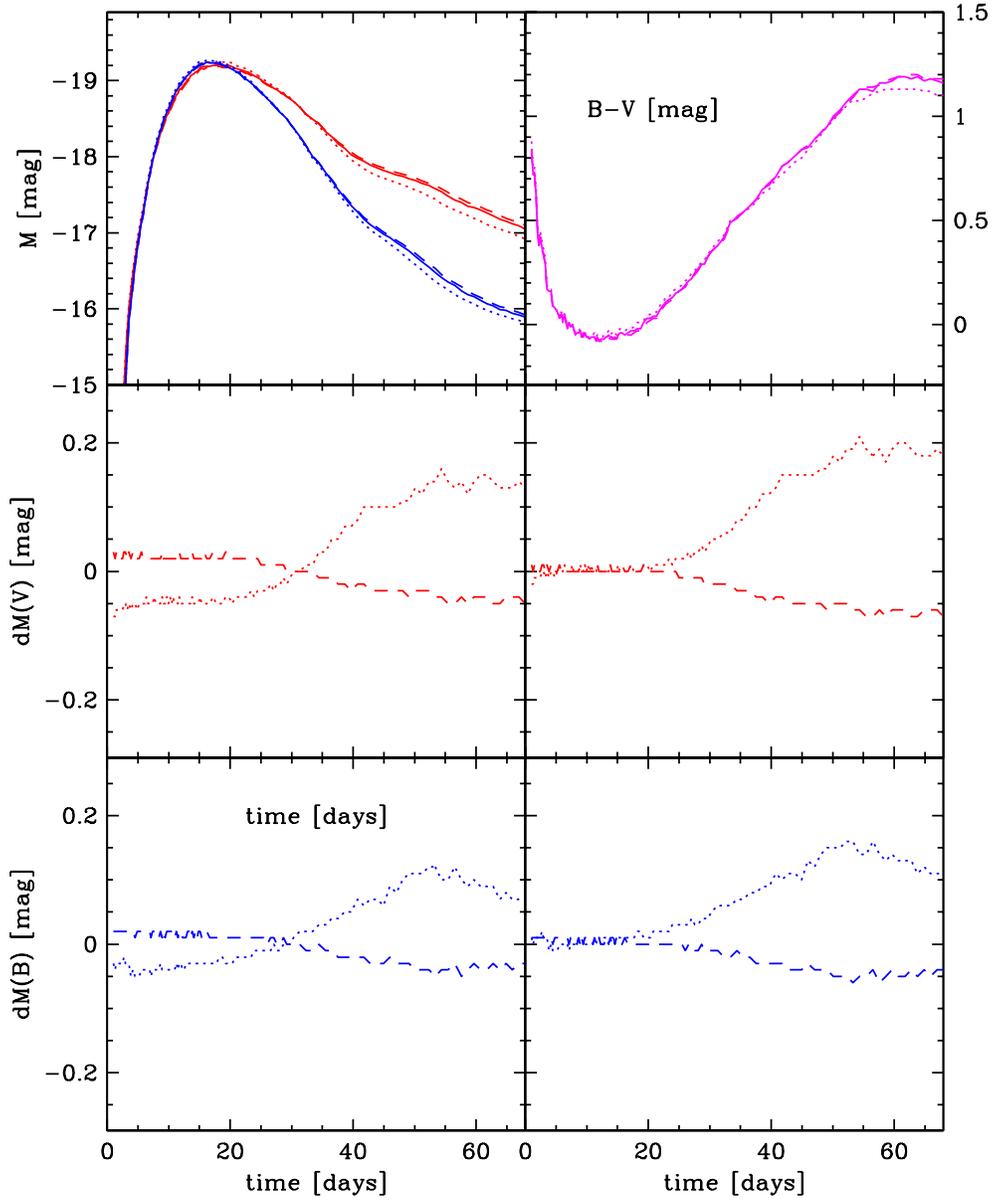}
\caption{Same as in Fig. \ref{prog} but compares models with fixed
  $\mms  = 5 M_\odot$ and $Z=0.02$ and with varying central densities 
 $\rho_c/(10^9$~\gcm)  of $1.5$ (dashed), $2.0$ (solid), and $6$
 (dotted). The central density is due to variations in the accretion
 rate \citep[see Fig.~6 of][]{h06}.
\label{rho}}
\end{figure}

Fig.~\ref{rho} shows the same type of comparison as Fig.~\ref{prog}
but for a series of models with fixed $\mms = 5M_\odot$ and solar
metallicity, and with varying accretion rate which leads to a  central
density $\rho_c = 1.5 \times 
10^9$, $2 \times 10^9$, and $6\times 10^9$~\gcm\ which, in our models,
corresponds to late-time accretion rates of $1\times 10^{-7} - 2
\times 10^{-6}$~$M_\odot$~yr$^{-1}$.  Fig.~\ref{rho} shows small
differences between the light curves prior to maximum light. They are
virtually unnoticeable in the upper panels of Fig.~\ref{rho} but can be
clearly seen on the differential plots (lower left panel). These
differences arise due to small variations in the binding energy of WD
models with different $\rho_c$.  When $\rho_c$ increases, so does the
binding energy, and this translates into a somewhat smaller expansion
velocity. As a result, light curves with larger/smaller $\rho_c$ rise
slower/faster and this results in a negative/positive pre-maximum
differential $dM$.

We can also see that variations in $\rho_c$ have a significant effect
on the behavior of light curves which begin to show up at $\simeq
20-25$ days after maximum light. Variations in $\rho_c$ lead to a
noticeable shift of the late time light curve with respect to the absolute
magnitude at maximum light. The light curves shift up or down when
$\rho_c$ decreases or increases, respectively.  This effect is related
to the existence of the central hole in the distribution of \nni\ in
Chandrasekhar-mass models of SNe~Ia.  Due to increasing electron
capture with density, the nuclear statistical equilibrium shifts away
from \nni\ to stable isotopes of the iron group when the central
density of a WD increases. 
 Near maximum and shortly past maximum
light the envelope of a SN~Ia is rather opaque and $\gamma$-rays
emitted near the center are trapped and do not contribute to the
formation of the light curve.  The light curve around maximum light is
controlled by the distribution of \nni\ in the outer parts of the
supernova.  As time goes on, the envelope expands and the distribution
of \nni\ near the center begins to influence the formation of the
light curve.  Larger central density means a larger hole and less
\nni. As a result, the light curve of a SN~Ia with higher $\rho_c$
becomes shifted down with respect to maximum. With decreasing $\rho_c$
the hole is smaller which means more \nni\ and the resulting shift of
the light curve is positive.  Note that ${B-V}$ is very similar but
at late times differs from the `general blue shift' produced by
variations in progenitors.  The lower right panels of Fig.~\ref{rho}
again show the differentials when a stretch correction 
 has been applied.  Small pre-maximum differences in the light
curves have virtually disappeared. However, shifts in the late light
curves caused by variations of \nni\ near the center remain very
pronounced.

Variations in some of the characteristics of light curves associated
with variations in $\rho_c$ and \mms\ are summarized in Table~1.
Table 1 shows that brightness of  SNe~Ia in both $M_V$ and $M_B$
increases monotonically from $-19.25$ to $-19.11$ and $-19.32$ to
$-19.18$ mag, respectively, when \mms\ decreases from $7$ to
$1.5M_\odot$ when other parameters, $\rho_{tr}$, $Z$, and $\rho_c$,
are kept constant. These variations are of the order $\simeq 0.1$~mag.
Variations of $M_V$ and $M_B$ with $\rho_c$ are smaller, $\pm 0.03$~mag,
because $\rho_c$ mostly affects the electron-capture in the center of
the SN~Ia which hardly contributes to the SN~Ia luminosity at maximum. The
effect of metallicity $Z$ can be seen by comparing the fiducial model
(first model) of Table 1 with $Z=0.02$ and the fifth model with
$Z=0.002$. $Z$ affects both $M_B$ and $M_V$ and has a pronounced
influence on the $B-V$ colors of SNe~Ia.

\begin{figure}
\epsscale{.80}
\includegraphics[width=.85\textwidth,angle=270]{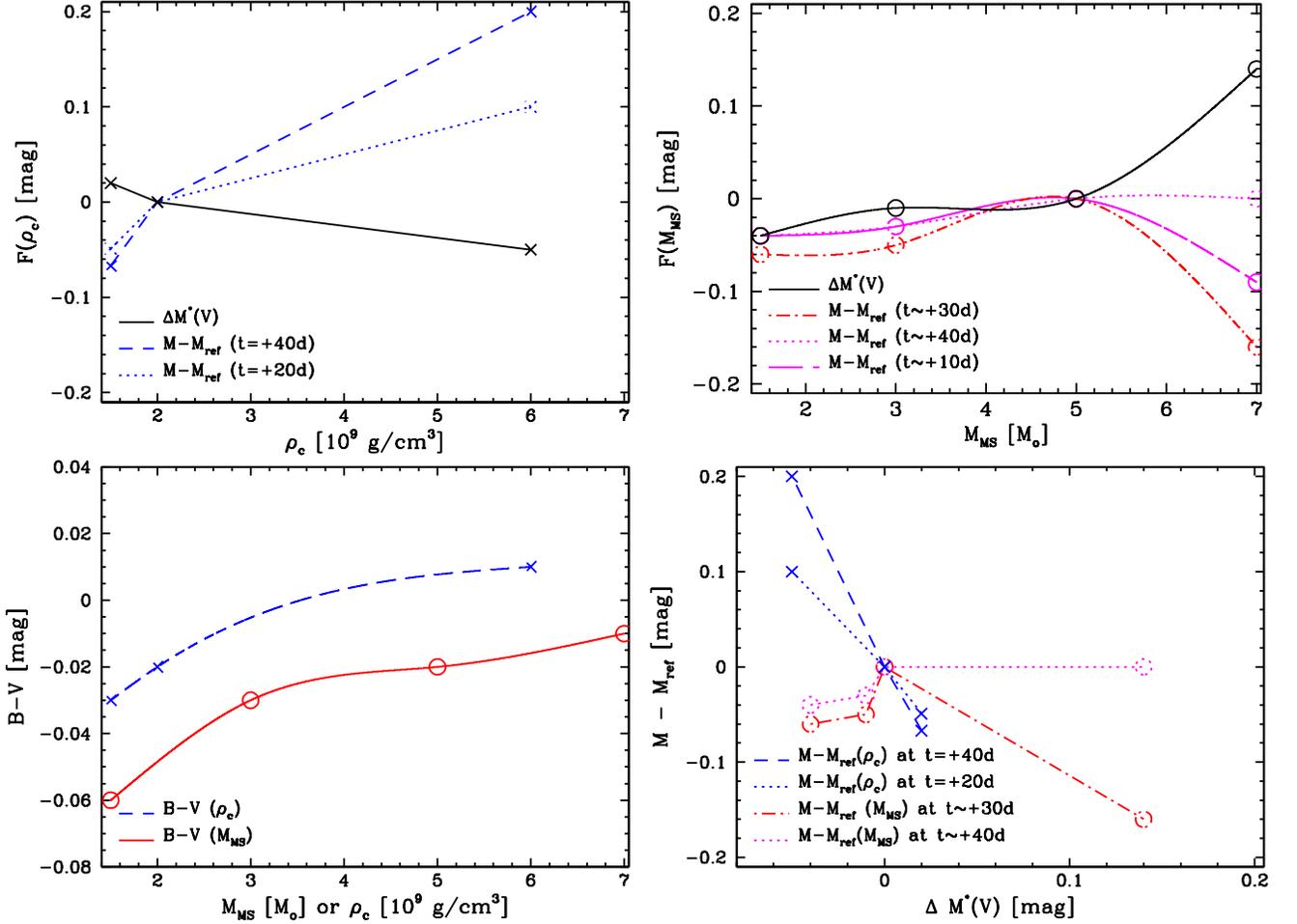}
\caption{
Basic observables in $V$ for models with varying central densities
$\rho_c $ (crosses) and main sequence progenitor masses \mms\ 
(open circles) relative to the reference model. All quantities are
normalized to the same fiducial $s$ factor using the brightness
decline relation \citep{h02}.
The change in absolute brightness at maximum light is $\Delta M^*(V)$,
the differential brightness is $M-\mref$ 
at times $+t$ after maximum light for the $\rho_c $-series (crosses, upper
left) and the $1^{st}$ ($\approx +7$~days), $2^{nd}$ ($\approx +35
$~days), and $3^{rd} +10 $~days 
for the \mms-series (open circles, upper right). The corresponding
colors $B-V$ (lower left) and $M-\mref$ ($\Delta M^*(V)$) (lower
right) are shown. 
Note that the sign of the residual between two SNe~Ia is arbitrary
depending on the choice  
of a reference object. 
\label{parameters}}
\end{figure}

Variations in $\rho_c$ and \mms\  have pronounced secondary 
\emph{differential} effects on stretch-matched light  
curves as  illustrated in Fig.~\ref{parameters}.  
(1) \mms\ influences the rise time of the light curves prior to maximum
light. Larger \mms\ leads to slower rise and vice versa.  The upper
right plot of Fig.~\ref{parameters} shows secondary variations of the
absolute visual magnitude of SNe~Ia, $M_V$, and differentials in
$M_V$ as a function of \mms\ for times $25$, $42$, and $55$ days past
maximum light.
(2) On the other hand, $\rho_c$ influences the light curve $\simeq 30$
days after maximum and later. Increases or decreases in $\rho_c$ cause
the later portion of the light curve to shift down or up with respect
to maximum, respectively.  The upper left plot of
Fig.~\ref{parameters} shows $M_V$, and differentials in $M_V$ as a
function of $\rho_c$ at $20$ and $40$ days past maximum light.
Secondary variations in $B-V$ for both series of models are given as
a function of \mms\  or $\rho_c$ in the lower left plot of
Fig.~\ref{parameters}. All the plots illustrate the point
that there are noticeable secondary variations of color and
absolute visual magnitude of stretch-matched supernovae which are
associated with variations of secondary parameters.

Finally, the lower right plot of Fig.~\ref{parameters} summarizes the
relation of a predicted relative variation in absolute brightness of
SNe~Ia (the quantity which cannot be determined from observations
unless the absolute brightness of SNe~Ia has been measured using an
independent method of calibration) and relative secondary variations
(differentials) of stretch-matched light curves of SNe~Ia which may be
directly accessible to observations.  This plot shows that secondary
variations in $M_V$ of stretch-matched supernovae may reach $0.2$ mag.

\section{Analysis of Observations}\label{Analysis}

The above theoretical considerations guide us 
in our differential analysis of observed light curves of SNe~Ia.

\begin{figure}
\epsscale{.80}
\plotone{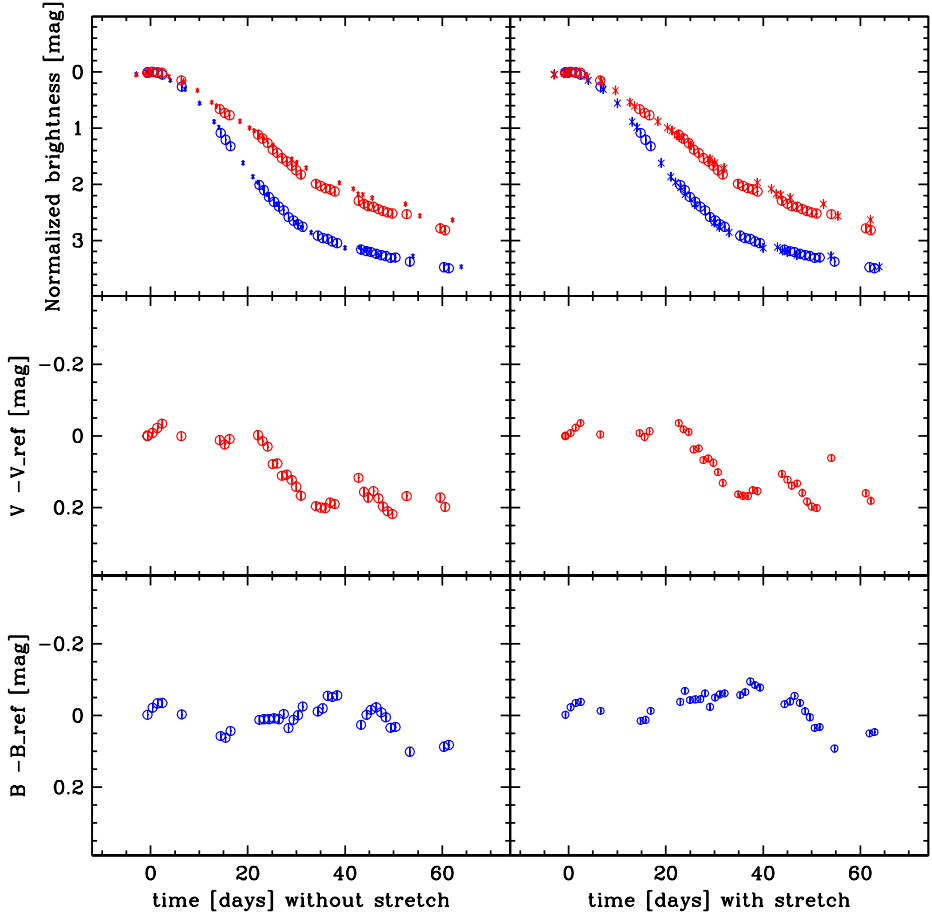}
\caption{Comparison of SN~2005al to SN~2005na chosen as a reference
  supernova. Left column - comparison of un-normalized
  supernovae. Right column -  
comparison of supernovae normalized to the maximum brightness in V and
with the $s$-factor of SN~2005al adjusted to be equal to the
$s$-factors of SN~2005na 
The original $s$-factors of all supernovae are listed in Table \ref{tbl-2}.
Upper row - B and V light curves. Middle row - differential in
V. Lower row - differential in B. 
\label{2005al_2005na}}
\end{figure}

\begin{figure}
\epsscale{.80}
\plotone{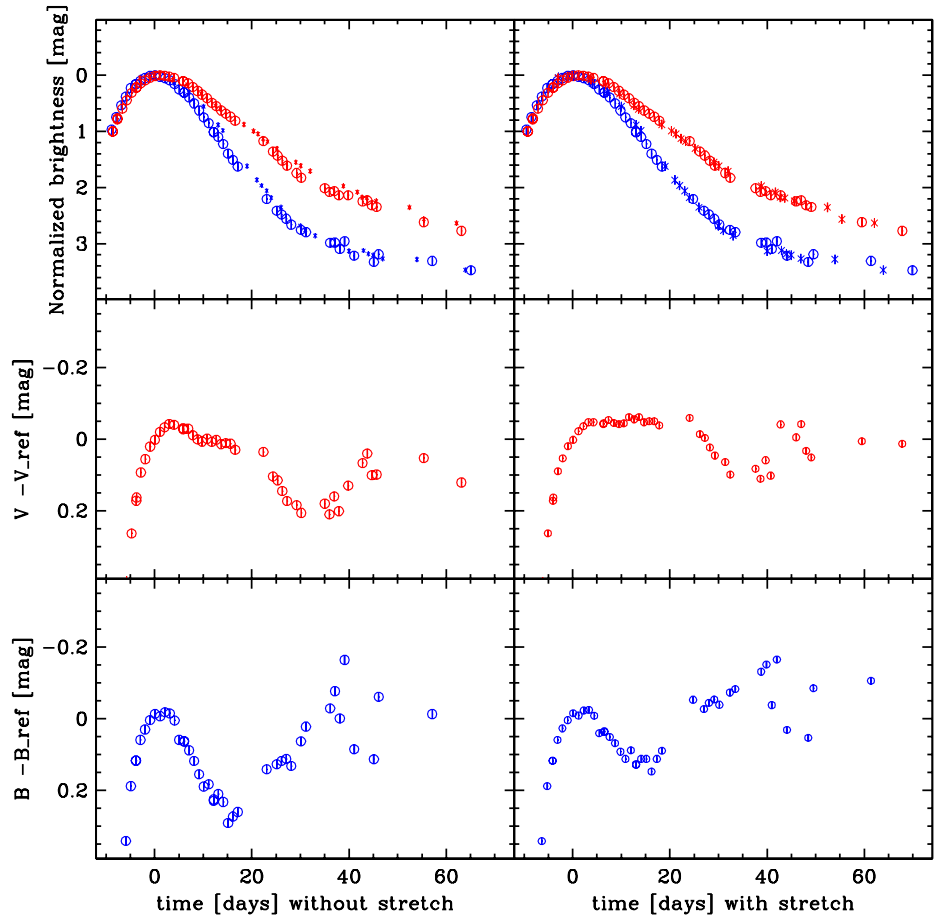}
\caption{Same as Fig.~\ref{2005al_2005na} but comparison of SN~2004ef
  to SN~2005na. 
\label{2004ef_2005na}}
\end{figure}

\begin{figure}
\epsscale{.80}
\plotone{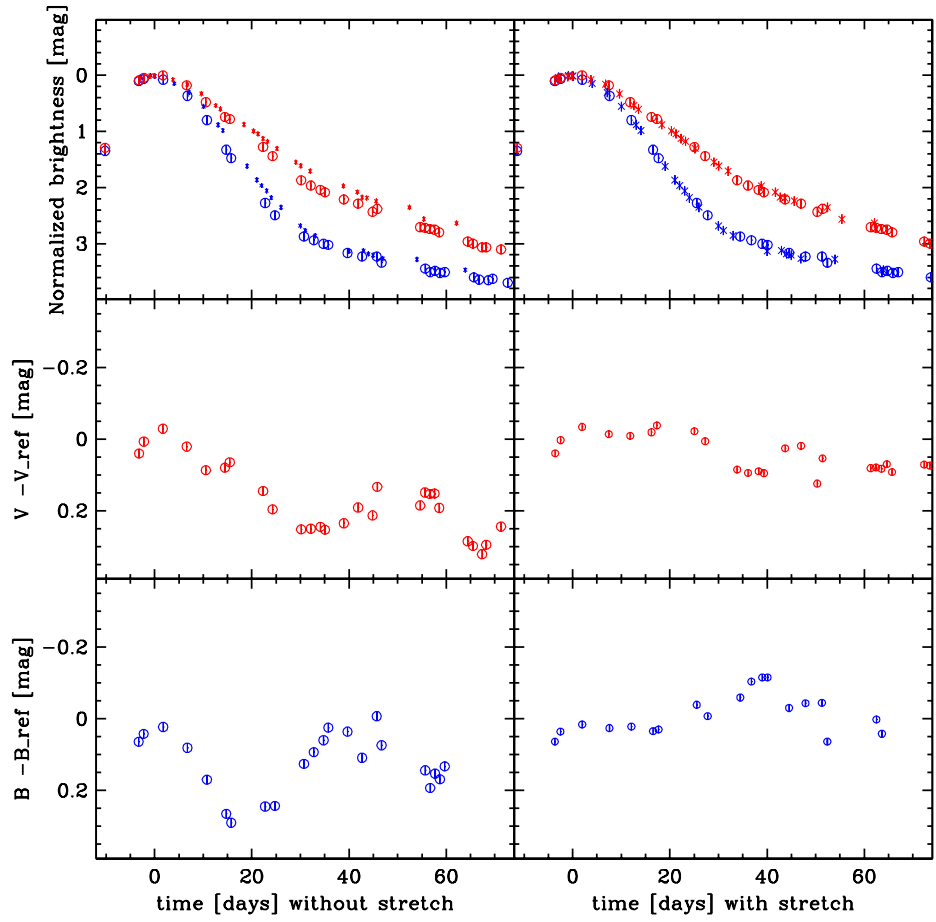}
\caption{Same as Fig.~\ref{2005al_2005na} but comparison of SN~2005ki
  to SN~2005na. 
\label{2005ki_2005na}}
\end{figure}

\begin{figure}
\epsscale{.80}
\plotone{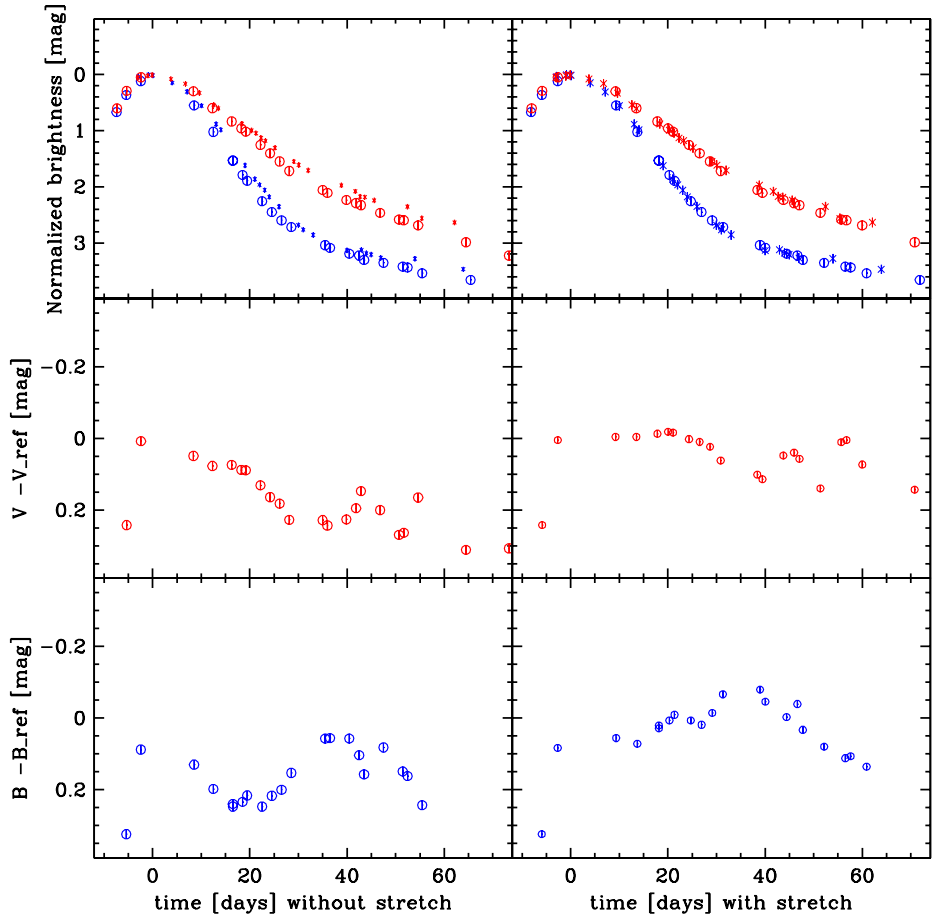}
\caption{Same as Fig.~\ref{2005al_2005na} but for SN~2005el. Note the
  S-shape in $M_-\mref$ with a `spread' of 
about $0.15$~mag.
\label{2004el_2005na}}
\end{figure}

\begin{figure}
\epsscale{.80}
\plotone{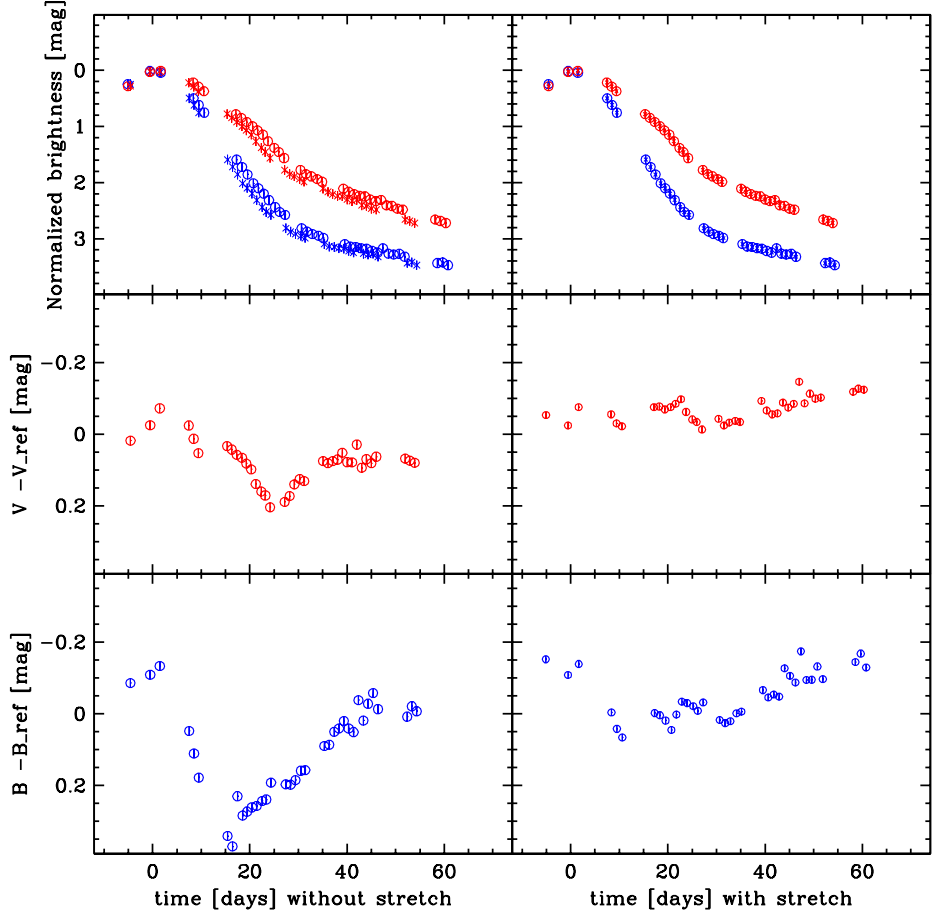}
\caption{Same as Fig.~\ref{2004el_2005na} but comparison of SN~2005am
  and  SN~2005el. Here, we used an offset in stretch $s$ 
and in $t_B$ of 0.01 and 0.5 days compared to Table 2,
respectively. The morphology of $M-\mref$ does not change, but the 
graph is `tilted' by about 0.1~mag.
\label{2004am_2005el}}
\end{figure}

Below we present a differential comparison of $V$ and $B$ light curves of
several SNe~Ia obtained by the CSP survey (Figs.~\ref{2005al_2005na} -
\ref{2004am_2005el}). The basic LC properties are given in Table
\ref{tbl-2}.  These objects were selected because they were discovered
well before maximum, have  good time coverage for up to $\simeq 60$~days
past maximum light, and very small observational errors.  
SN~2005na serves as the reference or fiducial model for the
comparison.  Stretch 
corrections for all SNe~Ia are listed in Table 1.  Left panels on each
figure show a comparison of light curves with stretch corrections from
Table 1.  Right panels show the comparison of SNe~Ia after we added an
additional stretch to their light curves in order to match the
$s$-factor of the fiducial model, SN~2005na. That is, rather than just
making the $\Delta m_{15}$ stretch correction using the formulae of
\citet{j06}, the curves were stretched such that the luminosity was
brought into coincidence with SN~2005na.  The additional stretch
is rather small since 
all these objects are in the range of normal bright SNe~Ia.  This
procedure helps reduce primary differences in due to the brightness
decline relation
(see \S~\ref{Theory}).

\noindent{SN~2005al vs.~SN~2005na} (Fig.~\ref{2005al_2005na}). These
supernovae are well within the normal bright range. Early portions
of their light curves, less than $25$ days after maximum light, are
very similar.  This is also indicated by the very small difference in
$s$-factors of these supernovae (Table~\ref{tbl-2}). Exact matching
gives only a marginal improvement in dispersion in $M - \mref$.
Fig.~\ref{2005al_2005na} 
shows that $M-\mref$ is less than a few hundredths of a
magnitude until $\simeq 25$ days after maximum. At later times the $V$
light curves are shifted with respect to each other by $\simeq 0.2$
mag.  There is no discernible systematic shift in $B$.  This behavior in
$V$ might be explained by variations in $\rho_c$ and the size of \nni\ hole
in the center of the SN~Ia.

\noindent{SN~2004ef vs.~SN~2005na} (Fig.~\ref{2004ef_2005na}). These
two supernovae are different in their pre-maximum behavior and show an
extremum between 20 and 40 days in $V$, both characteristics of variations
in the progenitor.  There might be a small systematic shift, $\simeq
0.05$~mag, in the late portions of the $V$ light curves as well, although it
is less pronounced compared to that of the previous  pair 
SN~2005al and {SN~2005na}. Differences between {SN~2004ef}
and {SN~2005na} may indicate both variations in progenitor masses and
metallicity,  and rather similar $\rho_c$ in these two events.

\noindent{SN~2005ki vs.~SN~2005na} (Fig.~\ref{2005ki_2005na}). Both
supernovae are very similar early on with a very small shift in the
late time light curves of less than $0.1$~mag.  They likely have similar
values of 
$\rho_c$, \mms,  and metallicity.

\noindent{SN~2005el vs.~SN~2005na} (Fig.~\ref{2004el_2005na}) These
two supernovae show differences in the early light curves, a secondary
extremum, and a significant shift in the late time light curves as
well.  Note 
that the stretches of SN~2005el and SN~2004ef are very similar, making
it not very likely that the huge $M$(SN~2005el)$-\mref$ can be
attributed to differences in $s$. In terms of SNe~Ia models, the
difference hints at variations in  progenitor mass \mms\  and 
$\rho_c$.

\noindent{SN~2005am vs.~SN~2005el (Fig. \ref{2004am_2005el})}.
Finally, we present a differential comparison of two supernovae with a
large difference in s-values.
 {SN~2005am}, is a
steep decliner with $\Delta m_{15}=1.61$ whereas {SN~2005el} is a normal
SNe~Ia with $\Delta m_{15}=1.37$. After stretching
(Fig. \ref{2004am_2005el}), the values of $M$(SN~2004am) and $\mref$
are surprisingly similar. Within the framework of models, this 
suggests similar progenitor masses and $\rho_c$.

Contrary to theoretical calculations which predict similar behavior of
$M-\mref$ in both $B$ and $V$, observations show, depending on the
example, variations in the $B$ and $V$ morphology.
At the moment the reason is not clear.  Several effects may play
a role: a) errors in $s$ and the time of maximum $t_{max}$, b)
differences in the theoretical and observational filter functions $B$
and $V$, and c) limitations of the explosion models and progenitors.

 First, we investigate the stability of differential comparisons of
light curves.  Matching the time of maximum light $t_{max}$ of two
supernovae may introduce an error in the differential
comparison. Another source of the differential error may be small
variations in $\Delta m_{15}$ or $s$.

\begin{figure}
\includegraphics[height=.85\textheight,angle=-90]{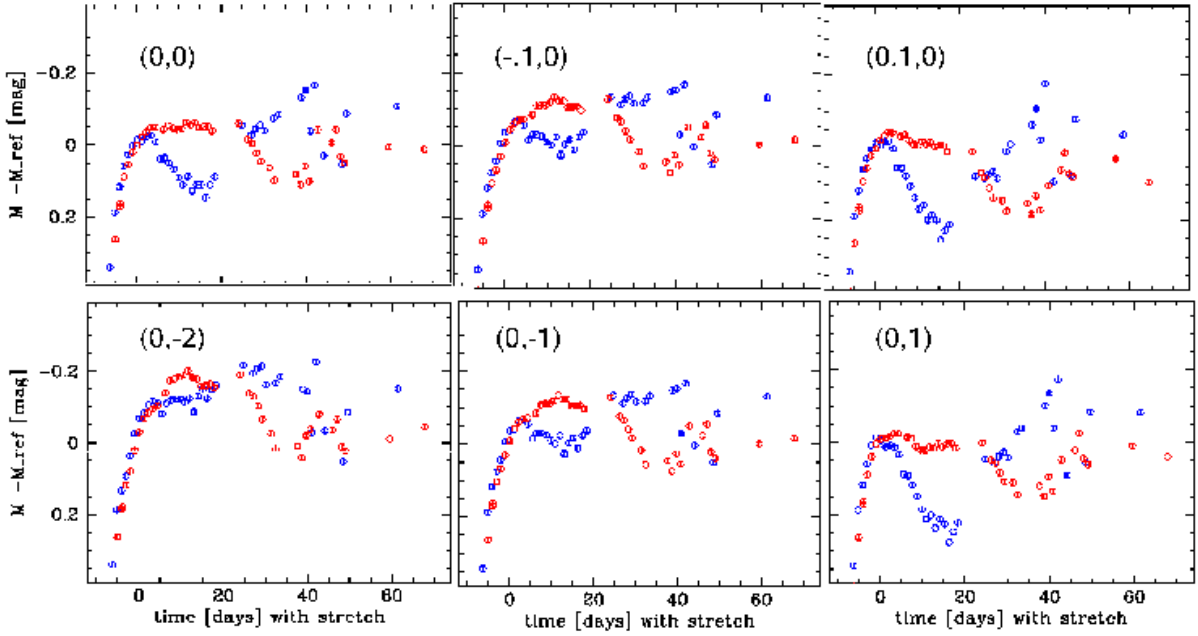}
\caption{Influence of uncertainties in $\Delta m_{15}$ and the time of
  maximum $t_{max}$ on the differential comparison of SN2004ef and
  SN2005na (Fig.~\ref{2004ef_2005na}). 
Panels are marked by the  assumed variations in  $\Delta
m_{15}$ and  $t_{max}$, $(\delta(\Delta m_{15}),\delta(
t_{max}))$. Time is in days. 
The functional form of the differential in $V$ appears to be stable including
the early rise, the extended plateau and the extremum at about 3 weeks
past maximum. The functional form of the differential in $B$ also
appears to be stable  
although quantitatively the differential is more sensitive to the
variations. 
\label{error}}
\end{figure}

Fig.~\ref{error} shows the effects of a relative shift of $t_{max}$
and a variation of $\Delta m_{15}$ of supernovae {SN~2004am} and
{SN~2004el}.  These two supernovae represent a pair with different
rise times and a corresponding deviation at $\simeq 30$ days after
maximum.  As discussed in \S~\ref{Theory} this may be attributed to 
variations in the progenitor masses of these two objects.  The differentials
between supernovae of various progenitors are more sensitive to
uncertainties in the observables than those with the $\rho_c$-signature
because the former show variations at both early and late times.

 Maxima were shifted by $\pm 1$ day and $2$ days, and $s$ by $\pm 0.1$.
The figure clearly shows that the morphology of $M-\mref$ in $V$ is stable,
namely the rise, a rather flat part around maximum, and a dip at day 30.
The variation of the differentials over the dip is less than $0.05$ mag.
For $B$, we still see that the main characteristics of $M-\mref$ in $B$
are stable, namely extrema at $\approx$ 15 and 40 days, but they are
larger than those in $V$ in size, e.g., the first dip varies 
by almost $0.3$ mag.

 The difference in the shapes of $B$ and $V$ differentials appear to
be real including the differences in $B$ between observed and
predicted morphology. In part, the size of the first $B$-dip may be
caused by fringing, i.e. shifts of LCs with several maxima and minima,
but hardly goes away.

 One possible effect is a wavelength offset in the $B$ filters.
Another group of effects includes
limitations inherent to the models: a) Metallicity variations for
stars with the same progenitor show differences of less than $0.05$
mag (Fig.~\ref{prog}), a value consistent with detailed spectral
analyses \citep{lentz01,h06}. However, uncertainties in our progenitor
evolution during central He-burning may underestimate the variations
in progenitors.  b) In $B$ at this level of accuracy, we may be seeing
limitations inherent in spherical models.  After maximum light, $B$ is
very sensitive to temperature variations because it is formed by the
Wien tail of the source function, $ S \propto e^{-h\nu/kT}$, where S
is the source function in the radiative transfer equation.  For
example, small variations of \nni\ mixing will increase the
temperature and, at the same time, the blocking in $B$. Indeed,
off-center DDT models show spectral changes in $B$ which are compatible
to the size of the deviation we find \citep{h06a,kasen09}.  However, strong
rotational mixing is rather unlikely because the impact on the
brightness decline relation and other observational constraints
discussed in \S~\ref{Introduction}.
 As seen above, uncertainties in the determination of $t_{max}$ and
 $s$ have moderate influence on the  
differential in $V$, but may amplify differences in $B$.

\section{Discussion and Conclusions}\label{DiscussionConclusions}

 We analyzed a set of high quality uniform $V$ and $B$
light curves of Type~Ia supernovae obtained by the Carnegie Supernova
Project. These data provide clear evidence for the existence of
secondary variations in SN~Ia light curves which are independent of the
primary Phillips relation and, thus, for the existence of at least two
independent secondary parameters. Comparison of the data with a series
of non-LTE light curve calculations of delayed detonation explosions
indicates that these secondary parameters may be physically related to
variations of central density, $\rho_c$ of a Chandrasekhar-mass white
dwarf exploding as a SN~Ia and to variations of the main sequence mass,
\mms, of the primary stellar companion in a progenitor binary
stellar system.

It is generally accepted that the total amount of \nni\ produced in
 the explosion is the primary parameter which controls the absolute
 brightness and the rate of post-maximum decline.  Production of
 nickel depends mainly on the transition density $\rho_{tr}$ at which
 subsonic burning turns into a supersonic detonation. However, if
 $\rho_{tr}$ were  the only parameter describing the explosion we
 would have a perfect one-dimensional sequence of SN~Ia events and a
 one-dimensional family of light curves. Some previous attempts to
 find additional parameters were done by comparing the risetimes of
 the nearby SNe~Ia sample to that of the distant one. \citet{riess99b}
 claimed to see a significant difference, however, \citet{goldhetal01}
 did not find the effect to be statistically significant.

The high quality light curves of SNe~Ia obtained by the CSP clearly
show that one parameter is insufficient to characterize the light
curve.  Light curves of several CSP 
supernovae shown in \S~\ref{Analysis} illustrate the fact that two
supernovae with identical behavior at maximum light and the early
portion of the post-maximum light curve, $0 - 30$ days past maximum,
may behave differently prior to maximum as well as at late times ($ >
30$ days past maximum light).  In particular, SNe~Ia that are similar
near maximum may have different pre-maximum rise and may have the late
portion of the $V$ light curve ($\geq 30$ past maximum light) shifted
up or down by different amounts with respect to maximum. Examples
presented in \S~\ref{Analysis} show that these differences may
reach $0.2 - 0.3$ mag.

Light curve modeling (\S~\ref{Theory}) predicts that two supernovae
with identical behavior of light curves near maximum may exhibit small
differences in the rise time to maximum and small shifts of the light
curves with respect to maximum luminosity at times $ \geq 30$ days
after maximum.  These variations may be traced to variations in
initial central density, $\rho_c$, and the WD progenitor mass on the
main sequence, \mms.  Instead of \mms, one may consider the
carbon-oxygen ratio C/O in the pre-exploding WD as a second parameter.
As far as light curves of SNe~Ia are concerned, these parameters,
$\rho_c$ and \mms\ (or C/O), may be treated as ``independent''
secondary parameters characterizing the light curve in addition to its
main characterization via $\Delta m_{15}$ or stretch.  Theoretical
predictions seem to be qualitatively consistent with two distinct
morphological types of deviations shown by observations in V
(\S~\ref{Analysis}). The B light curves also show distinct
morphological types of deviations but the morphology appears to be
different from that of the V light curves. On the contrary, light
curve modeling predicts similar morphological behavior in V and B.  We
note that most of the energy flux is coming in the V band and thus the
calculations of the V light curves should be more reliable.  The
reason for the discrepancy in B is not clear and requires further
investigation.

The existence of independent secondary parameters in SNe~Ia has several
 important implications. Obviously, any calibration of supernovae
 using a one-parameter set of light curve templates should lead to
 systematic errors in template matching and to calibration errors. The
 results of this paper indicate that a set of light-curve templates
 should form at least a three-parameter family. Our analysis suggests
 that a physically motivated set of templates may be constructed by
 using a primary SN~Ia template parameterized by $\Delta m_{15}$ or
 stretch with (1) an additional correction of the slope of the
 pre-maximum light curve and (2) an additional offset of the late V
 light curve with respect to maximum. Physically, the primary
 parameter $\Delta m_{15}$ should be thought of as a parameter which
 reflects variations in the amount and distribution of \nni\ in the outer
 parts of the SNe~Ia envelope responsible for the formation of the early
 post-maximum part of the light curve. The rise time correction might
 reflect the variation in the C/O ratio, and the offset of the light
 curve at later times might reflect variations in initial central density $\rho_c$
 and the amount and distribution of \nni\ in the central parts of the
 envelope.

However, producing a multi-parameter set of light-curve
templates is not sufficient for improving the calibration procedure.
The crucial point is that variations of initial conditions responsible
for secondary parameters must also cause, according to theoretical
predictions, small variations in absolute brightness of  SNe~Ia. The
calibration must take the dependence of the absolute brightness on
secondary parameters into account. By matching the proposed
multi-parameter templates with high-quality observations it should
be possible to account for secondary variations in the intrinsic
brightness and to reduce the calibration errors.  High quality data
are not sufficient at this time for carrying out such a program
systematically.  However, this work clearly indicates the 
potential of high quality uniform sets of observations for studying
secondary variations in SN~Ia light curves and for providing
important theoretical
 clues about physical mechanisms of such
variations.  One may hope that with improvements in SN~Ia theory,
verified with observations of nearby supernovae, it would be possible
to predict variations of absolute brightness of SNe~Ia as a function
of secondary parameters. This, in turn, may provide an opportunity to
predict evolutionary effects in SNe~Ia calibration related to
systematic changes in initial conditions with redshift.

Finally, we have to address the promise and limitations of this study.
Our quantitative theoretical predictions depend on our exact treatment
of the explosion mechanism and stellar evolution. Thus, variations in
the treatment of convection, the $^{12}C(\alpha, \gamma ) ^{16}O$
rate, rotation, and 3-D explosive effects could be important at some
level. Nevertheless we have tested the trends that we predict on a
small number of SNe~Ia and they should be tested on a large and
homogeneous set of SNe~Ia.  Our preliminary results are
promising. Light curves for a large number of SN are/will be available
by projects such as the ESSENCE \citep{essence}, CFH \citep{cfh}, NSNF
\citep{snfactory02spie}, PTF \citep{PTF09}, and LSST \citep{lsst}, and
we will publish an extensive comparison.  The properties of the
components, i.e. the shape in $M-\mref$, may be based on theoretical
models and optimized using large data sets.  However, even if our
results are confirmed fully, systematic theoretical studies including
3-D effects will be essential to go further.  As discussed in
\S~\ref{Introduction}, SN~Ia physics is intrinsically 3-D and thus
those effects must be taken into account. However, the last decade or
so of theoretical work indicates that the effect should not be
dominant since it appears from the the observations that 3-D effects
like rotation of the WD and the position of initial ignition are
reduced by the effects of the deflagration and DDT
\citep[e.g.]{khokhlov95,khokhlov97,Niemeyer99,R02,Gamezo03,Gamezo05,Plewa04,livne05,Roepke05}.
For a review see \citet{h06}. Finally, we return to the implications
of asymmetry for the use of SNe~Ia for cosmology.  A 10\% asymmetry of
the photosphere would not cause systematic difficulties for using SNe
Ia as distance indicators at the current level of accuracy of about
20\% \citep{wang03} though it would require that most of the
dispersion has to be attributed to asymmetry.  We note that if such
effects are present, then SNe~Ia are even more homogeneous than they
seem from current dispersions in peak brightness. This level of
asymmetry would, however, cause a directional dependence of the
luminosity of order $\sim 0.1$~mag \citep{h91} and a corresponding,
but smaller, dispersion in the brightness-decline relation of
SNe~Ia. This dispersion depends on the viewing angle dependence of the
luminosity variation and, thus, the nature of the asymmetry. The angle
dependence of the luminosity due to the viewing angle $\theta$ of a
single SN~Ia will not, in general, vary as the line of sight to the
equator as $cos\theta$. A more stringent limit comes from observations
of individual supernovae.  The first broad band survey by
\citet{wang96} established that SNe~Ia have very low polarization at a
level of $P\sim 0.2$\% whereas core-collapse SNe are generally more
highly polarized $P\sim 1$quality polarimetry measurements indicating that SNe~Ia are more
highly polarized before maximum, and a few weeks past maximum the
polarization disappears \citep{wang03}.  SN~2004dt showed that SNe~Ia
have spectral feature dependent polarization, implying different
chemical species have different geometry \citep{wang06,patat09}.
Since the continuum polarization at maximum light is observed to be $P
< 0.2 \%$ which, for scattering dominated atmospheres, translates into
a directional dependence of the flux at the 0.05 \% level
\citep{h91,wang97b,wang03,howell99by01,wang04,wang06,fesen07,patat09},
one expects that while 3-D effects are important for understanding the
explosion mechanism, their observational effects are relatively small.
However, as summarized in \citet{wwaraa08} asymmetries lead to a
dispersion in the color terms which can be magnified significantly
when extinction corrections are applied. Other independent factors
will contribute to the error and dispersion such that the interaction
within the progenitor system or the primordial metallicity require
early time spectra, and/or a combination of optical and IR data.

\acknowledgments We thank the anonymous referee for a very careful
reading of the manuscript and critial comments that much improved the
presentation. We thank Inma Dominguez for a careful reading of the
manuscript and helpful discussions and comments. The work presented in
this paper 
has been carried out 
within the NSF project ``Collaborative research:
Three-Dimensional Simulations of Type Ia Supernovae: Constraining
Models with Observations'' whose goal is is to test and constrain the
physics of supernovae by observations and improve SNe~Ia as tools for
high precision cosmology. The project involves The University of
Chicago (AST-0709181), the University of Oklahoma (AST-0707704),
Florida State University (AST-0708855), Texas A\&M (AST-0708873), The
University of Chile in Santiago, and the Las Campanas Observatory,
Chile.  This research was also supported, in part, by the NSF grant
AST-0703902 to PAH and US Department of Energy Award Number
DE-FG02-07ER41517 to EB. MH and GF acknowledge support from 
Fondecyt (1060808 and 3090004), 
Programa Iniciativa Cient\'ifica Milenio de MIDEPLAN' (P06-045-F), and
CONICYT (FONDAP 15010003 and PFB 06).
 The authors are especially grateful to the
members of the Carnegie Supernova Project team for the access to
observational data prior to publication.

\bibliographystyle{aa}  \bibliography{references}

\clearpage

\clearpage

\clearpage

\begin{deluxetable}{lcrrrrrr}
\tablecolumns{8}
\tablewidth{0pc}
\tablecaption{Properties of Calculated SNe~Ia \label{tbl-1}}
\tablehead{
\colhead{Parameter}  & \colhead{Model A}  &
    \colhead{Model B}  &
    \colhead{Model C} &
    \colhead{Model D}  &
    \colhead{Model E}  &
    \colhead{Model F}  &
    \colhead{Model G}  
}
\startdata
{$\mms$}       &{5.0}     &{1.5}   &{3.0}   &{7.0}   &{5.0}   &{5.0}   &{5.0}     \\
{$Z$}                    &{0.02}    &{0.02}  &{0.02}  &{0.02}  &{0.002} &{0.02 } &{0.02}     \\
{$\rho_c$}   &{2.0}     &{2.0}   &{2.0}   &{2.0}   &{2.0}   &{1.5}   &{6.0}     \\
$M_V$                      &  -19.21         & -19.25         &  -19.22         &  -19.11        &   -19.15       &    -19.19    &  -19.26 \\
$t_V$                 &   18.24         & 18.12          &   18.19         &  19.5          &   18.52        &      18.24      &   18.24   \\
$B-V$                      &   -0.02         &  -0.06         &   -0.03         &  -0.01         &   -0.13        &     0.03       &   0.01    \\
$M_B$                      &  -19.23         &  -19.32        &   -19.27        &  -19.18        &   -19.28       &      -19.22     &  -19.26    \\
$s/s_{f}$                  &   1.00          &   0.96         &    0.97         &  1.02          &   0.98         &      1.0          &   0.99  \\
\enddata
\tablecomments{Characteristics of theoretical light curves of DD
  models with various progenitor mass \mms/$M_\odot$, metallicity $Z$, and
  central density $\rho_c$ in units of $10^9$~\gcm.  Model A is the
  fiducial model with $\mms 
  = 5.0M_\odot$, $Z=0.02$ and $\rho_c = 2.0\times 10^9$~\gcm\ with
  $\Delta m_{15}=1.25$.  Listed for all models are the absolute
  maximum brightness $M_{B/V}$, time of V maximum $t_V$ (in days), and a
  correction to a stretch parameter, $s/s_f$, required for making the
  $s$-factor of the model equal to that of the fiducial model, $s_f$.}
\end{deluxetable}

\begin{deluxetable}{llllllllll}
\tablecolumns{10}
\tablewidth{0pc}
\tablecaption{Properties of observed SNe~Ia \label{tbl-2}}
\tablehead{
\colhead{SN}  & \colhead{$\Delta m_{15}$\tablenotemark{a}}  &
    \colhead{$s$\tablenotemark{b}}  &
    \colhead{$t_B/m_B$\tablenotemark{c}} &
    \colhead{$t_V/m_V$\tablenotemark{c}} & 
    \colhead{$\delta B({\rm r/s})$\tablenotemark{d}}    &
    \colhead{$\delta V({\rm r/s})$\tablenotemark{d}}          
}
\startdata
 2004ef & 1.47 &      &  264.96 & 264.54 &  0.09 /  0.05     &  0.00 / -0.03 \\
        & 1.33 & 0.88 &   16.92 &  17.06 &  0.03 / -0.07     &  0.14 /  0.04 \\
 2005al & 1.19 &      &  429.47 & 430.96 & -0.01 / -0.02       & -0.01 / -0.01 \\
        & 1.24 &0.92  &  15.08  & 15.08  &  0.01 / -0.03       &  0.16 /  0.13 \\
 2005am & 1.56 &      &  437.10 & 437.53 &  0.12 / -0.01       &  0.02 / -0.07 \\
        & 1.61 &0.75  &  13.76  &  13.84 &  0.14 / -0.09       &  0.29 / -0.06 \\
 2005el & 1.36 &      &  646.86 & 647.51 &  0.14 /  0.07       &  0.04 /  0.00 \\
        & 1.37 &0.86  &   15.24 & 15.22  &  0.13 /  0.02       &  0.24 /  0.08 \\
 2005ki & 1.44 &      &  705.98 & 706.20 &  0.12 /  0.03   &  0.03 / -0.01 \\
        & 1.41 &0.85  &   15.69 & 15.65  &  0.10 / -0.04   &  0.24 /  0.07 \\
 2005na & 1.19 &      &  740.32 & 741.79 &  0.00 /  0.00   &  0.00 /  0.00 \\
        & 1.19 &0.95  &   16.26 &  16.25 &  0.00 /  0.00   &  0.00 /  0.00 \\
\enddata
\tablenotetext{a}{Top  values of $\Delta m_{15}$ are the values for a family of $BV$ templates 
for which the B-band template gives the best fit to \textsl{all} $B$-band data.
Bottom values are derived from the ``early'' subset of B-data.
This subset includes data which extends to the inflection point, some
time $20-25$ days   
after maximum light, where the second derivative of $m_B$ changes sign.}

\tablenotetext{b}{$s$-factor is derived from the bottom value of $\Delta m_{15}$ (for 
the  early subset of B-data; see note (1)) 
using relation between $s$ and $\Delta m_{15}$ given in 
\cite{j06}.}

\tablenotetext{c}{Time (top number) and apparent maximum brightness
  (bottom number)  
at the truncated Julian Date is ``JD $-$ 2,453,000''.}

\tablenotetext{d}
{Average difference (in mag) between $B$ and $V$ light curves relative
to SN~2005el. 
Pairs of numbers are differences for raw (r) and stretched (s)
data. Top pair is determined  
using the early subset of data.  Bottom pair is determined using all
data. See note (a). }
\end{deluxetable}

\end{document}